\documentclass[runningheads]{CMSIM}

\usepackage{graphicx} % standard LaTeX graphics tool

\usepackage{amsmath,amssymb}

\usepackage{bm}
\usepackage{enumerate}
\allowdisplaybreaks
\newlength{\figfey}
\usepackage[usenames]{color} % for colour remarks

\usepackage{bm}
\def\eps{{\varepsilon}}

\newcommand{\fp}[1]{FP$ {\textrm{#1}}$}

\def\S{\mathcal{S}}

\def\eRM{{\mathrm e}}
\def\dRM{{\mathrm d}}

\def\mx{{\bm x}}

\def\mk{{\bm k}}

\def\eps{\varepsilon}

\allowdisplaybreaks
\allowbreak
\renewcommand{\thefootnote}

\title*{ 
 Stochastic Navier-Stokes equation for a compressible fluid: two-loop approximation
}

\titlerunning{\it Stochastic Navier-Stokes equation for a compressible fluid }

\author{
 Michal~Hnati\v{c}\inst{1,2,3}
 \and 
 Nikolay~M.~Gulitskiy\inst{4}
 \and 
 Tom\'a\v{s}~Lu\v{c}ivjansk\'y\inst{3} 
 \and
 Luk\'a\v{s}~Mi\v{z}i\v{s}in\inst{1,2}
 \and
 Viktor~\v{S}kult\'ety\inst{5}
}

\authorrunning{\it M.~Hnati\v{c} }

 \institute{
  Bogoliubov Laboratory of Theoretical Physics, 141980 Dubna, Russian Federation \\ 
 \and
  Institute of Experimental Physics, Slovak Academy of Sciences, Ko\v{s}ice, Slovakia \\ 
 \and
   Faculty of Sciences, \v{S}afarik University, Moyzesova 16, 040 01 Ko\v{s}ice, Slovakia \\   
 \and
 Department of Physics,  
 Saint-Petersburg State University, 7/9~Universitetskaya nab., St. Petersburg, 199034 Russian Federation \\
 \and 
  School of Physics and Astronomy, The University of Edinburgh, Peter Guthrie Tait Road, Edinburgh, EH9 3FD, United Kingdom\\
 (E-mail: {\tt hnatic@saske.sk, n.gulitskiy@spbu.ru, tomas.lucivjansky@upjs.sk,
 lukas.mizisin@gmail.com, viktoroslavs@gmail.com})  
 }

\begin{document}
\thispagestyle{empty}
\maketitle             
\setlength{\leftskip}{0pt}
\setlength{\headsep}{16pt}
\footnote{\begin{tabular}{p{11.2cm}r}
\small {\it $11^{th}$CHAOS Conference Proceedings, 5 - 8 June 2018, Rome, Italy} \\  
 %\small C. H. Skiadas (Ed)\\
   \small \textcopyright {} 2018 ISAST & \includegraphics[scale=0.38]{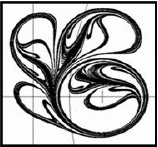}
 \end{tabular}}
\begin{abstract}
A model of fully developed turbulence of a compressible fluid is briefly reviewed. It is assumed that fluid
dynamics is governed by a stochastic version of Navier-Stokes equation. We show how
corresponding field theoretic-model can be obtained and further analyzed by means of
the perturbative renormalization group. Two fixed points of the RG equations
 are found. The perturbation theory is constructed
within formal expansion scheme in parameter $y$, which describes scaling behavior
of random force fluctuations. Actual calculations for fixed points' coordinates are performed
to two-loop order.
\end{abstract}

%---------------------------------------------------------------------------------------------------%
\section{ Introduction }\label{sec:intro}
%---------------------------------------------------------------------------------------------------%
Many natural phenomena are concerned with hydrodynamic flows. 
Ranging from microscopic up to macroscopic spatial scales 
 fluids can exist in profoundly different states. 
 Especially intrigued behavior is observed in case of turbulent flows. 
 Such flows are ubiquitous in nature and are more common than generally believed \cite{Frisch,Davidson}.
 Despite a substantial amount of effort that
 has been put into investigation of turbulence, the problem itself remains unsolved.
 
% efekt stlacitelnosti 
Most of studies are devoted to the case of incompressible fluid. However, 
 particularly in an astrophysical context we have to deal with a compressible fluid
rather than incompressible one \cite{Shore}. In recent years
there has also been an increased research activity of compressible turbulence in magnetohydrodynamic
context \cite{Carbone09,Sahra09,Eyink10,Galtier11,Banerjee16}.
In this work, our aim is to study compressible turbulence \cite{LL,Sagaut}, partially motivated by
  previous studies \cite{AK14,AGKL16,AGKL17_epj,AGKL17_pre}. 
 In case of a compressible medium, we are in fact
 examining system in which sound modes are generated. In fact, any compression leads to
acoustic (sound) waves that are transmitted through the medium and serve
as the prime source for dissipation. So the problem of the energy spectrum
(and dissipation rate) of a compressible fluid is essentially one of stochastic
acoustics. 

% RG pristup
%---------------------------------------------------------------------------------------------------%
The investigation of such behavior as anomalous scaling requires a lot of thorough analysis
to be carried out. The phenomenon manifests itself
in a singular power-like behavior of some statistical quantities (correlation functions, 
structure functions, etc.) in the
inertial-convective range in the fully developed turbulence regime~\cite{Frisch,Davidson,Falkovich2001}.
A quantitative parameter that describes intensity of turbulent motion is so-called 
Reynolds number $\mathrm{Re}$ that
represents a ratio between inertial and dissipative forces. For high enough values of 
$\mathrm{Re} \gg 1$
 inertial interval is exhibited in which just transfer of kinetic energy from outer $L$ (input)
  to microscopic $l$ (dissipative) scales take place. 
%-------------------------------------------------------------------------------%

A very useful and computationally effective approach to the problems with many interacting degrees of 
freedom on different scales is the field-theoretic 
renormalization group (RG) approach which can be subsequently accompanied by the operator product
expansion (OPE); see the monographs~\cite{Vasiliev,Amit,Zinn,turbo,Tauber}. 
One of the greatest challenges is an investigation of the Navier-Stokes equation for a compressible 
fluid, and, in particular, 
 a passive scalar field advection by this velocity ensemble. The first relevant discussion and analysis 
 of passive advection emerged a few 
decades ago for the Kraichnan's velocity ensemble~\cite{Kraich1,GK,RG,ant06}. Further studies developed
 its more realistic 
generalizations~\cite{AG12,J13,Uni3,VectorN,JJ15,V96-mod,amodel,HHL}. The RG+OPE technique 
 was also applied to more complicated models, in 
particular, to the 
compressible case~\cite{AK14,NSpass,Ant04,AGM,JJR16,VM,tracer2,tracer3,ANU97,AK15,AK2,LM,St,MTY}.
 Our aim here is to improve existing (one-loop) results on compressible stochastic Navier-Stokes equation
 and determine relevant physical quantities to two-loop order. Note that in contrast
 to static phenomena transition from one-loop
 to two-loop approximation pose in stochastic dynamics much more demanding task. 
 
%-------------------------------------------------------------------------------%
%-------------------------------------------------------------------------------%

The paper is a continuation of previous works~\cite{AGKL16,AGKL17_epj,AGKL17_pre} and it is
organized as follows. In the 
introductory Sec.~\ref{sec:model} 
we give a brief overview of the model and we reformulate stochastic equations
into field-theoretical language.
 Sec.~\ref{sec:RG} is devoted to the renormalization group analysis.
In Sec.~\ref{sec:scaling} we present
 the fixed points' structure, describe possible scaling regimes and calculate
  critical dimensions. 
 The concluding Sec.~\ref{sec:conclusion} is devoted to a short discussion and future plans.
%-------------------------------------------------------------------------------%

%---------------------------------------------------------------------------------------------------%
\section{ Model }\label{sec:model}
%---------------------------------------------------------------------------------------------------%

Let us start with a discussion of a model for compressible velocity fluctuations. 
The dynamics of a compressible fluid is governed by the stochastic Navier-Stokes equation~\cite{LL}
taken in the form
\begin{equation} 
  \rho\nabla_{t} v_{i} = \nu_{0} [\delta_{ik}\partial^{2}-
  \partial_{i}\partial_{k}] v_{k}
  + \mu_0 \partial_{i}\partial_{k} v_{k} - \partial_{i} p + f^v_{i},
  \label{eq:NS}
\end{equation}
where the operator $\nabla_t$ stands for  an expression $\nabla_{t} = \partial_{t} + v_{k} \partial_{k}$, 
also known as a Lagrangian (or convective) derivative. Further,  
$\rho=\rho(t,\mx)$ is a fluid density field, $v_i=v_i(t,\mx)$ is the velocity field,
$\partial_{t} =
\partial /\partial t$ is a time derivative, $\partial_{i} = \partial /\partial x_{i}$ is a $i$th component
of spatial gradient, 
$\partial^{2} =\partial_{i}\partial_{i}$ is the Laplace operator, $p=p(t,\mx)$ is the pressure field,  
and $f^v_i$ is the external force, which is specified later. 
In what follows
 we employ a condensed notation in which we write  $x=(t,{\bm x})$, where a spatial
  vector variable $\mx$ equals $(x_1,x_2,\ldots,x_d)$ with $d$ being a dimensionality of space.
   Although it is possible to consider $d$ as additional free parameter \cite{AGKL17_pre}, in this work
   spatial dimension $d$ implicitly takes most physically relevant value $3$.
Two parameters $\nu_{0}$ and $\mu_{0}$ in Eq.~(\ref{eq:NS}) are two
 viscosity coefficients~\cite{LL}. Summations over repeated vector indices (Einstein summation
 convention) are always implied in this work.

 Let us make two important remarks regarding the physical
 interpretation of Eq.~(\ref{eq:NS}). First, this equation should 
 be regarded as an dynamic equation only for a fluctuating part of the total velocity field. In other words,
 it is assumed that the mean (regular) part of the velocity field has already been subtracted 
 \cite{Frisch,Davidson}.
 Second, the random force $f_i^v$ mimics not only an input of energy, but to some extent
 it is responsible for neglected
 interactions between fluctuating part of the velocity field and the mean part \cite{Vasiliev,turbo}. 
 In reality, the latter interactions are always present and their mutual interplay
 generates turbulence \cite{Davidson}. 
 
 Let us note that stochastic theory of turbulence
 is similar to a fluctuation theory for critical phenomena \cite{Vasiliev,papo}. The main difference
 is lack of Hamilton-like operator for turbulence. Nevertheless, it is still possible to take advantege of
  well-established theoretical tools borrowed from quantum field theory and employ them
 on turbulence \cite{Vasiliev,Zinn}.
 
To complete the theoretical set-up of the model, Eq.~(\ref{eq:NS}) 
has to be augmented by additional two relations. They are a continuity equation and
 a certain thermodynamic relation \cite{LL}.
 The former one can be written in the form
\begin{equation}
  \partial_{t} \rho  + \partial_{i} (\rho v_{i})   = 0
  \label{eq:CE}
\end{equation}
and the latter {we choose as} follows
\begin{equation}
  \delta p = c_0^2 \delta\rho,
  \label{eq:SE}
\end{equation}
where $\delta p$ and $\delta \rho$ describe deviations from the equilibrium values of pressure field
 and density field, respectively.

Viscous terms in Eq.~(\ref{eq:NS}) characterize
dissipative processes in the system and in a turbulent state it is expected their relevance
 at small length scales. Without a continuous input of energy,
 turbulent processes would eventually die out because of dissipation and {the flow would eventually
  become} regular.
 There are various possibilities for modeling of energy input \cite{turbo}. For translationally
 invariant theories it is convenient to specify properties of the random force $f_i$ in 
 time-momentum representation
\begin{equation}
  \langle f_i(x) f_j(x') = \frac{\delta(t-t')}{(2\pi)^d} \int_{k>m} \dRM^d k \mbox{ }
  D^v_{ij}({\mk})
  \eRM^{i{\mk}\cdot({\mx-\mx'})},
\label{eq:ff}
\end{equation}
where the delta function in time variable ensures Galilean invariance of the model \cite{turbo}. The integral
 in Eq.~(\ref{eq:ff}) is  infrared~(IR) regularized with a parameter $m\sim L_v^{-1}$, where 
$L_v$ denotes outer scale, i.e., scale of the biggest turbulent eddies. More
details can be found in the literature \cite{turbo,JETP}. 
 The kernel function ${D}^v_{ij}({\bm k})$ is now assumed in the following form
\begin{equation}  
  D_{ij}^v({\bm k})=g_{0} \nu_0^3 k^{4-d-y} \biggl\{
  P_{ij}({\bm k}) + \alpha Q_{ij}({\bm k})
  \biggl\},
  \label{eq:correl2}
\end{equation}
where $g_{0}$ is a coupling constant, $k=|\mk|$ is the wave number, $y$ is a suitable scaling
exponent, and $\alpha$ is a free dimensionless parameter. Parameter $\alpha$ basically measures
intensity with which energy flows into a system via longitudinal modes. 

 Further, the projection operators $P_{ij}$ and $Q_{ij}$
 in the momentum space read
\begin{equation}
   P_{ij} (\mk) = \delta_{ij} - \frac{k_i k_j}{k^2}, \quad Q_{ij} = \frac{k_i k_j}{k^2},
   \label{eq:project}
\end{equation}
and correspond to the transversal and longitudinal projector, respectively.

Due to its functional form with respect to momentum dependence, 
 function (\ref{eq:correl2}) corresponds to a non-local term in ensuing field theoretic
 action.  However, physical and plausible mathematical considerations \cite{Vasiliev} justify
 this choice. One of the reasons is a straightforward modeling of
 a steady input of energy into the system from outer scales. In what follows we attack the problem
 with the RG approach. The value of the scaling exponent $y$ in Eq.~(\ref{eq:correl2})
 describes a deviation from a
logarithmic behavior (that is obtained for $y=0$).
In the stochastic theory of turbulence the main interest is in the limit behavior
 $y\rightarrow 4$ that yields an ideal pumping from infinite spatial scales \cite{turbo}.

 Let us make a brief remark about possible generalization of the model. Although, we
 present our results with a general spatial dimension $d$, we have always implicitly in mind its
 most realistic value $d=3$. However, it would be possible to generalize the model \cite{AGKL17_pre}
 and consider $d$ as additional small parameter, similar to the well-known $\varphi^4-$theory
 in critical statics \cite{Zinn,Tauber}. 
Usually the spatial dimension $d$ plays a passive role and
  is considered only as an independent parameter. However, Honkonen and Nalimov \cite{HN96}
  showed that in the vicinity of space dimension {$d=2$} additional divergences appear in the model
  of the incompressible Navier-Stokes ensemble and
  these divergences have to be properly taken into account. Their procedure also naturally leads
  into improved perturbation expansion \cite{AHKV05,AHH10}.
    As can be seen from the RG discussion in the next section a 
    similar situation occurs for the model (\ref{eq:NS})
   in the vicinity of space dimension $d=4$. In this case an additional divergence
   appears in the 1-irreducible Green function $\left\langle v'v'\right\rangle_{\text{1-ir}}$. 
   Utilizing this feature  one can employ a double expansion scheme, in which the 
formal expansion parameters are $y$, and $\eps=4-d$, i.e., a deviation from
the space dimension $d=4$ \cite{HHL,HN96}. 
 
Our main theoretical tool is the renormalization group theory. Its proper application requires
 a proof of a renormalizability of the model, i.e., a proof that only a finite number 
 of  divergent structures exists in a diagrammatic expansion \cite{Amit,Zinn}.
 As was shown in \cite{VN96}, this requirement can be accomplished 
 by the following procedure:
 first the  stochastic equation~(\ref{eq:NS}) is divided by density field $\rho$, 
 then fluctuations in viscous terms are neglected, and finally.
 Using the expressions~(\ref{eq:CE}) and~(\ref{eq:SE})
the problem is formulated into a system of two coupled differential equations
\begin{align}
  \nabla_{t} v_{i} & = 
  \nu_{0} [\delta_{ik}\partial^{2}-\partial_{i}\partial_{k}]
  v_{k}\! +\! \mu_0 \partial_{i}\partial_{k} v_{k} -\!
  \partial_{i} \phi\! +\! f_{i},
  \label{eq:ANU} \\
  \nabla_{t} \phi & =  -c_{0}^{2} \partial_{i}v_{i},
  \label{eq:ANU1}
\end{align}
where a new field $\phi=\phi(x)$ has been introduced for convenience. 
It is related to the density fluctuations via the 
relation $\phi = c_0^2 \ln (\rho/\overline{\rho})$ \cite{AK14,ANU97}.
A parameter $c_0$ denotes the adiabatic speed of sound, $\overline{\rho}$ is the mean value of 
 density field $\rho$, and 
$f_{i}=f_{i}(x)$ is the external force normalized per unit mass.

According to the general theorem~\cite{Vasiliev,Zinn}, the stochastic problem given by Eqs.
 (\ref{eq:ANU}), and (\ref{eq:ANU1}),
is tantamount to the field theoretic
model with a doubled set of fields $\Phi=\left\{v_{i}, v_{i}',\phi, \phi'\right\}$ and given
 De Dominicis-Janssen action 
functional. The latter can be written
in a compact form as a sum of two terms 
\begin{align}
  \S_\text{total} [\Phi] & = \S_\text{vel}[\Phi] + \S_\text{den}[\Phi], 
  \label{eq:full_action} 
\end{align}
where the first term describes a velocity part
\begin{align}
  \S_\text{vel}[\Phi] & = \frac{v_i' {D}_{ij}^v v_j'}{2} 
  +v_i' \biggl[
  -\nabla_t v_i + \nu_0(\delta_{ij}\partial^2 - \partial_i \partial_j)v_j
  +u_0 \nu_0 \partial_i \partial_j v_j - \partial_i \phi
  \biggl],
  \label{eq:vel_action1}
\end{align}
and the second term is given by the expression
\begin{align}
  \S_\text{den}[\Phi] = \phi'[-\nabla_t \phi  + v_0 \nu_0 \partial^2 \phi - c_0^2 (\partial_i v_i)].
  \label{eq:vel_action2}
\end{align}  
Here, $ {D}^v_{ij}$ is the correlation function~(\ref{eq:correl2}). Note that we have introduced 
a new dimensionless parameter
$u_0=\mu_0/\nu_0>0$ and a new term
$v_0 \nu_{0} \phi' \partial^{2}\phi$ with another positive
dimensionless parameter $v_0$, which is needed  to ensure 
multiplicative renormalizability \cite{Vasiliev,Zinn}. 

 Further, we employ a condensed notation, in which integrals over the spatial variable 
${\mx}$ and the time variable $t$, as well as summation over repeated indices, are not
 explicitly written, 
for instance
\begin{align}  
  {\phi'}\partial_t{\phi} & =\int\! \dRM t \!\int \dRM^d{x}\, \phi'(t,\mx)\partial_t\phi(t,\mx), \nonumber \\
  v'_iD_{ik}{v'}_k
  & = \sum_{ik}\int\! \dRM t \!\int\! \dRM^d x \!\! \int\! \dRM^d x' 
  \,v_i(t,{\mx})D^v_{ik}({\mx}-{\mx'})v_k(t,{\mx'}). 
  \label{eq:quadlocal2}
\end{align}

In a functional formulation various stochastic
quantities (correlation and structure functions) are calculated as path
integrals with weight functional 
\begin{equation*}
  \exp (S_\text{total}[\Phi]). 
\end{equation*}
 The main benefits of such approach are transparency in a perturbation theory and
 potential use of  powerful methods
of the quantum field theory, such as Feynman diagrammatic technique and 
renormalization group procedure \cite{Zinn,turbo,Tauber}.
 
\begin{figure}
  \centerline{
    \includegraphics[width=0.9\textwidth]{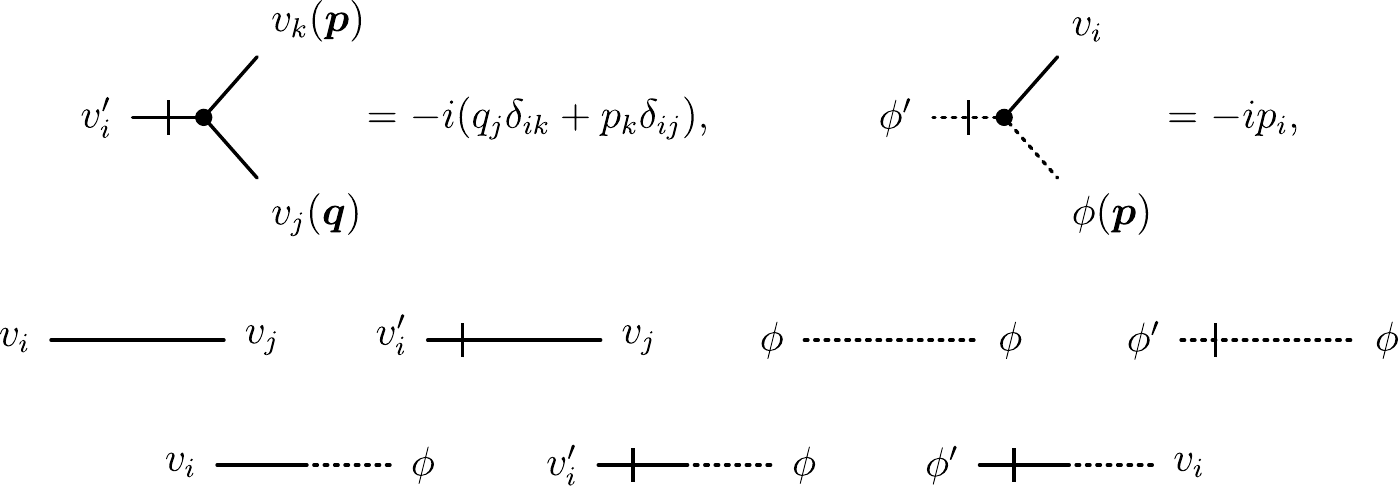}}
  \caption{Graphical representation of elements of the perturbation theory 
   (\ref{eq:full_action}). }
  \label{fig:pert}
\end{figure}

%---------------------------------------------------------------------------------------------------%
\section{ Renormalization group analysis }\label{sec:RG}
%---------------------------------------------------------------------------------------------------%

Ultraviolet renormalizability reveals itself in a presence divergences 
in Feynman graphs, which are constructed according to simple laws \cite{Vasiliev,Tauber} using
 a graphical notation from Fig. \ref{fig:pert}.
From a practical point of view,
 an analysis of the 1-particle irreducible Green functions, later referred to as 1-irreducible Green functions
 following the notation in~\cite{Vasiliev}, is of utmost importance. 
In the case of translationally invariant models \cite{Vasiliev,Tauber}
two independent scales have to be introduced: the time scale $T$ and the length scale $L$. Thus
the canonical dimension of any quantity $F$ (a field or a parameter) is
described by two numbers, the
frequency dimension $d_{F}^{\omega}$ and the momentum dimension $d_{F}^{k}$,
defined such that following normalization holds
\begin{equation}
  d_k^k =-d_{ x}^k=1,\quad d_k^{\omega} =d_{x}^{\omega }=0,\quad
  d_{\omega }^{\omega }=-d_t^{\omega }=1,\quad d_{\omega }^k=d_t^k=0,
  \label{eq:def_dim}
\end{equation}
% \begin{align}
%   d_k^k&=-d_{ x}^k=1, &d_k^{\omega}& =d_{x}^{\omega }=0,\nonumber\\
%   d_{\omega }^{\omega }&=-d_t^{\omega }=1, &d_{\omega }^k&=d_t^k=0,
%   \label{eq:def_dim}
% \end{align}
and the given quantity then scales as
\begin{eqnarray}
[F] \sim [T]^{-d_{F}^{\omega}} [L]^{-d_{F}^{k}}.
\label{eq:canon}
\end{eqnarray}
The remaining dimensions can be found from the requirement that each term of the
action functional (\ref{eq:full_action}) be dimensionless, with respect to both the momentum and the
frequency dimensions separately.

Based on $d_F^k$ and $d_F^\omega$  the total canonical dimension
$d_F=d_F^k+2d_F^\omega$ can be introduced, which in
the renormalization theory of dynamic models plays the
same role as the conventional (momentum) dimension does in
static problems \cite{Vasiliev}. 
 Setting $\omega \sim k^{2}$ ensures that all the viscosity and diffusion coefficients in the model are dimensionless. 
Another option is to set the speed of sound $c_{0}$
dimensionless and  consequently obtain that $\omega \sim k$, i.e., $d_{F}=d_{F}^{k}+d_{F}^{\omega}$.
This variant would mean that
we are interested in the asymptotic behavior of the Green functions
as $\omega \sim k \to 0$, in other words, in sound modes in turbulent medium. Even though this problem
is very interesting itself, it is not yet accessible {for} the
RG treatment, so we do not discuss it here.
The choice $\omega \sim k^{2} \to 0$ is the same as in the models 
of incompressible fluid, where
it is the only possibility because the speed of sound is infinite.
A similar alternative in dispersion laws exists, for example, within the so-called model H of equilibrium
dynamical critical behavior, see~\cite{Vasiliev,Tauber}.

The canonical dimensions for  the model~(\ref{eq:full_action}) are listed in
Tab.~\ref{tab:vel}. It then directly follows that the model
is logarithmic (the coupling constant $g \sim [L]^{-y}$
becomes dimensionless) at $y=0$.
In this work we use the minimal subtraction (MS) scheme for the calculation
of renormalization constants. In this scheme the UV divergences in the
Green functions manifest
themselves as pole in $y$
\begin{table*}
\centering
\caption{Canonical dimensions of the fields and parameters entering velocity part of the 
total action (\ref{eq:full_action}).}
\label{tab:vel}
  \begin{tabular}{c|c|c|c|c|c|c|c|c|c}
    $F$ & $ v_i'$ & $ v_i$ & $\phi'$ & $\phi$  &
    $m$, $\mu$, $\Lambda$ & $\nu_0$, $\nu$ & $c_{0}$, $c$ &
    $g_{10}$  & $u_{0}$, $v_{0}$ $w_{0}$, $u$, $v$, $g$, $\alpha$  \\
    \hline
    $d_{F}^{k}$ & $d+1$ & $-1$ & $d+2$ & $-2$  
    & 1 &  $-2$ & $-1$ & $y$  & 0 \\
    $d_{F}^{\omega}$ & $-1$ & 1 & $-2$ & 2 & 
    0 & 1 & 1 & 0  & 0\\
    $d_{F}$ & $d-1$ & 1 & $d-2$ & 2 &  1 & 0 & 1 & $y$  & 0 \\
  \end{tabular}
\end{table*}

The total canonical dimension of any 1-irreducible Green function $\Gamma$ is given by the relation
\begin{eqnarray}
\delta_{\Gamma} = d+2 - \sum_{\Phi} N_{\Phi} d_{\Phi},
\label{index}
\end{eqnarray}
where $N_{\Phi}$ is the number of the given type of field {entering the function}
$\Gamma$, $d_{\Phi}$ is the corresponding total canonical dimension of field $\Phi$, and
the summation runs
over all types of the fields $\Phi$ in function $\Gamma$~\cite{Vasiliev,Zinn,Tauber}.  

Superficial UV divergences whose removal requires counterterms can be present only in
those functions $\Gamma$ 
for which
the formal index of divergence $\delta_{\Gamma}$ is a non-negative integer.
A dimensional analysis should be augmented by the several additional considerations.
 They are all explicitly stated in the previous works \cite{AK14,AGKL17_pre}. Therefore, we do not repeat them here
and continue with a simple conclusion that model with the action (\ref{eq:full_action}) is
renormalizable. 

From a straightforward inspection of RG theory it is clear that for determination of
scaling regimes only two Green functions have to be considered. The reason is that we study
theory with three charges, $g,u$ and $v$. Once their fixed values are found, we would be able
to study scaling regimes and their stabilities. 
  Thus only graphs that are needed to be calculated are
two-point Green functions $\langle v v\rangle_\text{1PI}$ and $\langle p p \rangle_\text{1PI}$. 
In a one-loop approximation \cite{AK14,AGKL17_pre,ANU97} the calculation is simple 
as there are only two Feynman diagrams
at this level
\begin{align} 
  &  \raisebox{-1.ex}{ \includegraphics[width=1.8cm]{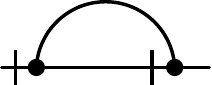}} 
     \raisebox{-1ex}{ \includegraphics[width=1.8cm]{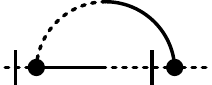}}     
     \label{eq:one_loop}.
\end{align}
For two-loop approximation, following graphs have to be computed for the velocity part
\begin{align} 
  &  \raisebox{-1ex}{ \includegraphics[width=\figfey]{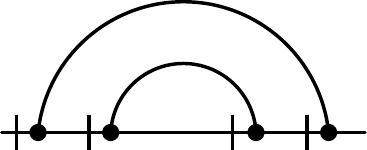}}
     \raisebox{-1ex}{ \includegraphics[width=\figfey]{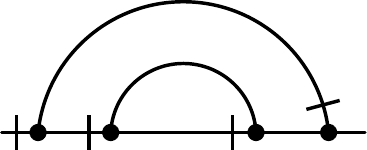}}     
     \raisebox{-1ex}{ \includegraphics[width=\figfey]{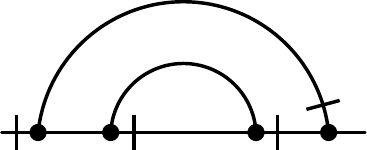}}  
     \raisebox{-1ex}{ \includegraphics[width=\figfey]{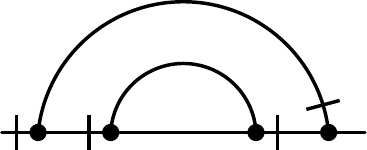}}
     \nonumber \\  &    
     \raisebox{-1ex}{ \includegraphics[width=\figfey]{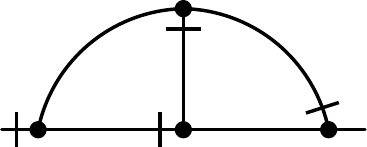}}     
     \raisebox{-1ex}{ \includegraphics[width=\figfey]{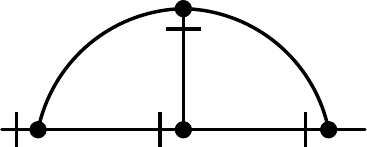}}     
     \raisebox{-1ex}{ \includegraphics[width=\figfey]{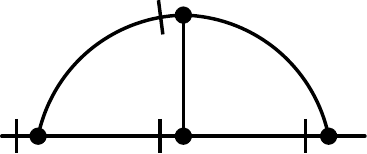}}     
     \raisebox{-1ex}{ \includegraphics[width=\figfey]{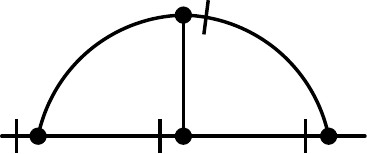}}     
     \label{eq:vel_grafy1}.
\end{align}
On the other hand, for the pressure part additional eight diagrams are needed
\begin{align} 
  &  \raisebox{-1ex}{ \includegraphics[width=\figfey]{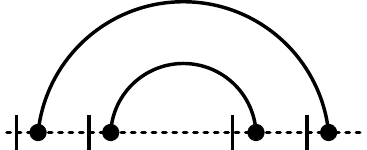}}
     \raisebox{-1ex}{ \includegraphics[width=\figfey]{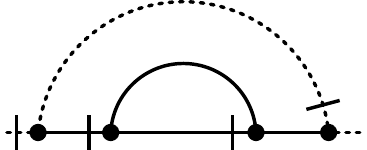}}     
     \raisebox{-1ex}{ \includegraphics[width=\figfey]{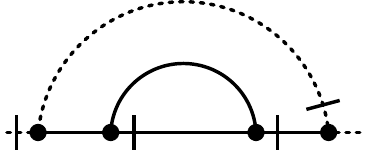}}  
     \raisebox{-1ex}{ \includegraphics[width=\figfey]{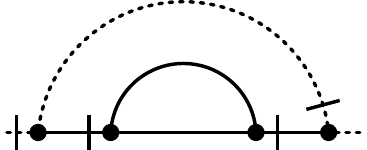}}     
     \nonumber \\  &
     \raisebox{-1ex}{ \includegraphics[width=\figfey]{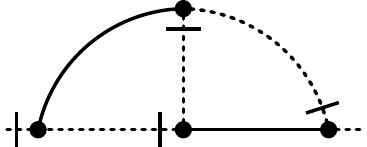}}     
     \raisebox{-1ex}{ \includegraphics[width=\figfey]{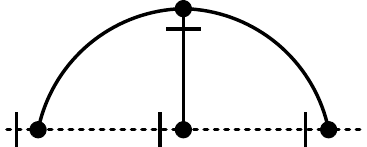}}     
     \raisebox{-1ex}{ \includegraphics[width=\figfey]{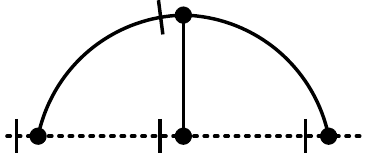}}     
     \raisebox{-1ex}{ \includegraphics[width=\figfey]{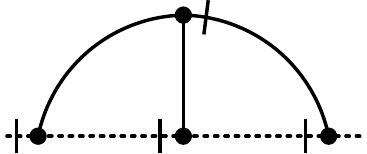}}     
     \label{eq:vel_grafy2}.
\end{align}
The remaining diagrams are needed only for determination of anomalous dimension
of fields, which is left for future study. 

In contrast to the incompressible case \cite{AHKV05} compressible model (\ref{eq:full_action}) proved
to be much more demanding from technical point of view. This is caused by three reasons. First, in compressible
case there are six physical quantities ($\mu_0,\nu_0,v_0,g_0,\alpha,c_0$) instead of just two ($\nu_0$ and charge $g_0$)
 for incompressible fluid.
Second, propagators now contain both transversal and longitudinal parts and last, interaction vertices are not
proportional to the momentum of prime field, what implies that the degree of UV divergence could not be lowered.

In evaluation of UV divergent parts of Feynman diagrams we have applied approach suggested in \cite{AHKV05}.
 Using symbolic software \cite{wolfram} we were able to simplify some calculations and to determine
 divergent parts at least in numerical sense.
Because the details of calculation are rather straightforward
and proceed in a standard fashion \cite{Vasiliev,Amit,Zinn,Tauber}, we refrain from
mentioning them here.

%---------------------------------------------------------------------------------------------------%
\section{ Scaling regimes  }\label{sec:scaling}
%---------------------------------------------------------------------------------------------------%

The relation between the initial and renormalized action functionals 
$\S(\varphi,e_{0})= \S^{R}(Z_\varphi\varphi,e,\mu)$ (where $e_{0}$
is the complete set of bare parameters and $e$ is the set of their renormalized
counterparts) yields the fundamental RG differential equation:
\begin{equation}
\biggl\{ {\cal D}_{RG} + N_{\varphi}\gamma_{\varphi} +
N_{\varphi'}\gamma_{\varphi'} \biggr\} \,G^{R}(e,\mu,\dots) = 0,
\label{RG1}
\end{equation}
where $G =\langle \varphi\cdots\varphi\rangle$ is a correlation function of the fields~$\varphi$;
$N_{\varphi}$ and $N_{\varphi'}$ are the counts of normalization-requiring
fields $\varphi$ and $\varphi'$, respectively, which are the inputs to
$G$; 
the ellipsis in expression~\eqref{RG1} stands for the other arguments of $G$ (spatial
and time variables, etc.).
${\cal D}_{RG}$ is the operation $\widetilde{\cal D}_{\mu}$
expressed in the renormalized variables and
$\widetilde{\cal D}_{\mu}$
is the differential operation $\mu\partial_{\mu}$ for fixed
$e_{0}$. For the present model it takes the form
\begin{equation}
{\cal D}_{RG}= {\cal D}_{\mu} + \beta_{g}\partial_{g}  +
\beta_{u}\partial_{u} + \beta_{v}\partial_{v}
- \gamma_{\nu}{\cal D}_{\nu}- \gamma_{c}{\cal D}_{c} .
\label{RG2}
\end{equation}
Here, we have denoted ${\cal D}_{x} \equiv x\partial_{x}$ for any variable $x$.
The anomalous dimension $\gamma_{F}$ of some quantity $F$
(a field or a parameter) is defined as
\begin{equation}
\gamma_{F}= Z_{F}^{-1} \widetilde{\cal D}_{\mu} Z_{F} =
\widetilde{\cal D}_{\mu} \ln Z_F ,
\label{RGF1}
\end{equation}
and the $\beta$ functions for the four dimensionless coupling
constants $g$, $u$ and $v$, which we now redefine according to the following rule
\begin{equation}
   g \equiv g_1, \quad u \equiv g_2, \quad v \equiv g_3.
\end{equation}
for convenience.
$\beta$ functions express the flows of parameters under
the RG transformation, and are defined through relation $\beta_{i} = \widetilde{\cal D}_{\mu} g_i$. 
This yields
\begin{align}
  \beta_{1} & =  g_1\,(-y-\gamma_{1}), \quad 
  \beta_{2}  =  -g_2\gamma_{2}, \quad
  \beta_{3}  = -3\gamma_{3}, \\
  \gamma_1 & \equiv \gamma_g, \quad 
  \gamma_2 \equiv \gamma_u, \quad
  \gamma_3 \equiv \gamma_v.
  \label{eq:all_beta}
\end{align}
 Based on the analysis of the RG equation~\eqref{RG1} it follows that 
the large scale behavior with respect to spatial and time scales is
governed by the IR attractive (``stable'') fixed points $g^*\equiv\{g_1^*,g_2^*,g_3^*\}$,
 {whose} coordinates are found from the conditions~\cite{Vasiliev,Amit,Zinn}:
\begin{align}
  &\beta_{1} (g^{*}) = \beta_{2}
  (g^{*}) = \beta_{3} (g^{*}) = 0.
  \label{eq:gen_beta}
\end{align}
 Let us consider a set of invariant couplings $\overline{g}_i = \overline{g}_i(s,\{g_i\})$
with the initial data $\overline{g}_i|_{s=1} = g_i$. Here, $s=k/\mu$ 
and IR asymptotic behavior (i.e., behavior at large distances) corresponds
to the limit $s\rightarrow 0$. An evolution of invariant couplings is described by
the set of flow equations
\begin{equation}
  \mathcal{D}_s \overline{g}_i = \beta_i(\overline{g}_j),
  \label{eq:invariant_chrg}
\end{equation}
whose solution as $s\to0$ behaves approximately like
\begin{equation}
  \overline{g}_i(s,g^*) \cong g_i^*+const\times s^{\omega_i},
  \label{Asym}
\end{equation}
where $\left\{\omega_i\right\}$ is the set of eigenvalues of the matrix 
\begin{equation}
\Omega_{ij}=\partial\beta_{i}/\partial g_{j}|_{g^{*}}.
\label{Omega}
\end{equation}
The existence of IR {attractive} solutions of the RG equations leads
to the existence of the scaling behavior of Green functions. 
From~\eqref{Asym} it follows that the type of the fixed point is determined by the matrix~\eqref{Omega}:
for the IR {attractive} fixed points the matrix $\Omega$ has to be positive definite.

 Altogether two IR attractive fixed points are found, which defines possible scaling regimes of the system.
The fixed point \fp{I} (the trivial or Gaussian point) is stable if $y<0$. This regime is characterized
by irrelevance of all his charges, i.e.,
\begin{equation}
  g_1^* = g_2^* = g_3^* = 0.
  \label{fp1}
\end{equation}
On the other hand, the fixed point \fp{II} is fully nontrivial, i.e. all his coordinates attain
 nonzero value. We have found the following numerical expressions for them
\begin{align}
  g_1^* & = 2y + \frac{-2.00625\alpha^2-4.8847\alpha+4.4206}{5\alpha+12}y^2,\\
  g_2^* & = 1 + \frac{0.125797\alpha^2 - 0.83854\alpha -0.188233}{5\alpha+12}y,\\
  g_3^* &= 1 + \frac{0.217295\alpha^3 + 1.7247\alpha^2-1.27116\alpha-6.9228}{(\alpha+6)(5\alpha+12)}y.
  \label{eq:fp2}
\end{align}
To one-loop order we have thus obtained same results as has been claimed previously
\cite{AK14,ANU97}. The initial analysis reveals that \fp{II} is nontrivial for $y>0$ and not
very large values of $\alpha$.

%---------------------------------------------------------------------------------------------------------------
\section{Conclusion}\label{sec:conclusion}
%---------------------------------------------------------------------------------------------------------------

In the present paper the compressible fluid governed by the Navier-Stokes velocity
ensemble has been examined. The fluid was assumed to be
compressible and the space dimension was fixed to $d=3$. The problem has been investigated by means of 
renormalization group and expansion scheme in $y$ was constructed. 

There are two nontrivial IR stable fixed points in this model and, therefore, the critical behavior in the inertial 
range demonstrates two different regimes depending on the 
 the scaling exponent $y$. Coordinates of nontrivial fixed points have been obtained for the first time
 to two-loop precision. This can be considered as a first step to full two-loop analysis of the model.

%------------------------------------------------------------------------------------------------%
\section*{Acknowledgements}
The work was supported by VEGA grant No.~1/0345/17 of the Ministry
of Education, Science, Research and Sport of the Slovak Republic,
 {and  by the Russian Foundation
for Basic Research within the Project No. 16-32-00086}.
N.~M.~G. acknowledges the support from the Saint Petersburg Committee of Science and High School.

%------------------------------------------------------------------------------------------------%

\end{document}